\documentclass{ws-ijmpd}
\usepackage[super]{cite}
\usepackage{xcolor}
\usepackage[verbose,hypertexnames=false]{hyperref}
\hypersetup{colorlinks=false,allbordercolors=blue,pdfborderstyle={/S/U/W 1}}

\usepackage{float} 
\usepackage{mathtools}
\usepackage{slashed}
\usepackage[retainorgcmds]{IEEEtrantools}

\begin{document}

\markboth{George Georgiou \& Dimitrios Zoakos}
{Holographic interpolations of codimension-2 defect CFTs}

%
\catchline{}{}{}{}{}
%

\title{Holographic interpolations of codimension-2 defect CFTs}

\author{George Georgiou}

\address{Agios Georgios Pagon, 49083, 
Corfu, Greece\\
georgios.georgiou2@gmail.com}

\author{Dimitrios Zoakos}

\address{Department of Physics, 
University of Patras, 26504 Patras, Greece\\
dzoakos@upatras.gr}

\maketitle


\begin{abstract}
We provide a comprehensive overview of the current status of higher-codimension defect systems. We review the holographic description and field theoretic properties of codimension-2 defects within the framework of defect Conformal Field Theories (dCFTs). Starting from the well-established classification of $1/2$-BPS supersymmetric defects, we examine their realisation through probe branes and bubbling supergravity geometries. Special emphasis is placed on recent developments involving non-supersymmetric D3/D5 configurations and their holographic interpolations. We discuss the calculation of important physical observables, such as one-point functions of the stress-energy tensor and chiral primary operators, across both weak and strong coupling regimes. The agreement of the results in the two regimes exhibits the full power of the holographic principle.\\
This is a proceedings contribution to the Athens Workshop in Theoretical Physics: 10th Anniversary, held at the National and Kapodistrian University of Athens on December 17-19 2025.\\

\end{abstract}



\section{Introduction}

A wide range of physical phenomena, spanning from boundaries and interfaces to extended objects like branes, strings, and Wilson lines, can be effectively modeled through Conformal Field Theory (CFT) dynamics in the presence of defects.
The introduction of a defect partially or completely breaks the conformal symmetry of the ambient CFT, a process that significantly increases the complexity of calculating physical observables. 
Within the dCFT framework, several key characteristics define the system, such as the amount of preserved supersymmetry (SUSY), the integrability of the associated boundary conditions, and the co-dimension of the defect itself. 

In the context of the gauge/gravity duality, defects are holographically represented by extended objects in the bulk, typically probe branes wrapping appropriate submanifolds of the \(AdS_5 \times S^5\) geometry. 
The prototypical example is the codimension-1 defect, exemplified by the 
D3/D5 brane intersection, where a stack of D3-branes ends on a D5-brane, preserving 
\(\mathcal{N}=(4,4)\) supersymmetry. 
The holographic dual is realized in Ref.~ \refcite{Karch:2000gx} 
by a probe D5-brane wrapping an \(AdS_4 \times S^2\) 
subspace within \(AdS_5 \times S^5\). 
Natural holographic duals for defects are generally branes whose worldvolume contains an \(AdS\) factor, which encodes the conformal symmetry on the defect. 
For the codimension-2 case, a primary example is the Gukov-Witten defect. 
Its holographic description involves a probe D3-brane wrapping an \(AdS_3 \times S^1\) subspace, a configuration that preserves 1/2 of the original supersymmetry. 

On the field theory side, these codimension-2 defects are characterized by classical solutions where certain scalar fields acquire non-zero diagonal vacuum expectation values (vevs). 
These vevs exhibit a simple pole on the defect, with their spacetime dependence strictly dictated by conformal invariance. 

While the holographic description of codimension-2 defects is often associated with probe D3-branes, as seen in the Gukov-Witten type, recent developments have extended this landscape. Specifically, the dCFT frameworks analyzed in Refs.~\refcite{Georgiou:2025mgg} and ~\refcite{Georgiou:2025wbg} explore generivally non-supersymmetric codimension-2 defects realised via D5-brane configurations. In these models, the probe branes wrap appropriate internal cycles that preserve the conformal symmetry on the defect, while the field theory side is characterized by scalar vevs with the characteristic pole structure. This expansion of the known codimension-2 examples is crucial for testing the universality of the gauge/gravity duality in set ups with less symmetry involving higher-codimension systems.

The present review is organized as follows: In Sec.~\ref{sec:GW}, we provide a brief overview of the codimension-2 $1/2$-BPS configurations, examining their complementary descriptions in gauge theory, probe brane dynamics, and bubbling supergravity geometries. Sec.~\ref{Holo_interpolations} is dedicated to the study of holographic interpolations of dCFTs, with a particular focus on the novel D5-brane solutions that connect supersymmetric and non-supersymmetric regimes. Within this framework, we review the computation of one-point functions for the stress-energy tensor and chiral primary operators, demonstrating the agreement between weak and strong coupling results in the appropriate limits. Finally, in Sec.~\ref{conclusions}, we summarize our findings and discuss recent progress regarding the defect Weyl anomaly coefficients $b$ and $d_1$.


\section{The Gukov-Witten 1/2 BPS defect}
\label{sec:GW}

Before embarking on the study of the non-supersymmetric, holographically 
realised dCFTs of Refs.~\refcite{Georgiou:2025mgg} and \refcite{Georgiou:2025wbg} 
let's us very briefly review the co-dimension two ${1/2}$-BPS configurations 
considered in  Refs.~\refcite{Gomis:2007fi} and \refcite{Drukker:2008wr}.

Extended operators play a central role in gauge theory. In 4-dimensional ${\cal N} = 4$ supersymmetric Yang–Mills (SYM) theory, familiar examples include Wilson and ’t Hooft loop operators,
which probe electric and magnetic features of the theory.
Surface operators in Ref.~\refcite{Gukov:2006jk} generalise these constructions to codimension-two defects supported on a two-dimensional surface $\Sigma \subset \mathbb{R}^4$. They provide a richer class of observables and are particularly important for:
\begin{itemize}
\item understanding S-duality,
\item probing non-perturbative dynamics,
\item establishing connections with geometric representation theory.
\end{itemize}
Supersymmetry and/or conformal symmetry greatly simplify the structure and properties of those operators.
A key insight is that surface operators admit three equivalent descriptions: as singular configurations in gauge theory, as probe branes in string theory, and as smooth ``bubbling'' supergravity geometries in Ref.~\refcite{Gomis:2007fi}. 
In what follows, although we will not elaborate on the bullet points above, we will briefly visit the three descriptions one after the other.


\subsection{Gauge theory}\label{GT}

In generic gauge theory with gauge group $G=U(N)$ a surface operator ${\cal O}_\Sigma$
located on a two-dimensional surface $\Sigma$ embedded in $\mathbb{R}^{1,3}$ spacetime is partly characterised by a vortex configuration for the gauge field $A$ which has the form
\begin{eqnarray}\label{AA}
   A= \left(\begin{array}{cccc} \alpha_{1}\otimes 1_{N_1 \times N_1} & 0 & \cdots &0 \\ 0 &\alpha_{2}\otimes1_{N_2 \times N_2}&\cdots&0\\ 
\vdots &\vdots &\ddots&\vdots\\
0&0&\cdots&\alpha_{M}\otimes1_{N_M \times N_M} \end{array}\right)_{N\times N}\, d\psi
=\alpha_{N} \, d\psi\,. 
\end{eqnarray}
The gauge field \eqref{AA} induces a co-dimension two singularity supported on the defect $\Sigma$.
Indeed, the corresponding field strength is then given by 
\begin{eqnarray}
    F=dA=2 \pi \alpha_N\delta_\Sigma  \,\,\,  {\rm where}  \,\,\,  \delta_\Sigma=d(d\psi)
\end{eqnarray}
Here, $\delta_\Sigma$ is a two-form delta function supported on the surface of the operator. Furthermore, $\psi$ is the polar angle in the ${\mathbb R}^2 \subset {\mathbb R}^{1,3}$ plane normal to $\Sigma$.\footnote{In this section, we will take $\Sigma$ to be either a flat 2-dimensional plane or an $S^2$.} The surface operator ${\cal O}_\Sigma$
 is additionally characterised by a set of 2-dimensional $\theta$-angles along $\Sigma$ which introduce the following operator insertion into the ${\cal N}=4$ SYM path integral:
\begin{equation}
  \exp\left(i\sum_{l=1}^M\eta_l\int_\Sigma\hbox{Tr}\;{F_l}\right).  
\end{equation}
From \eqref{AA} it is obvious that the gauge group $G=U(N)$ is broken down to the Levi group $L = \prod_{l=1}^M U(N_l)$. Therefore, the matrix of $\theta$-angles is given by the $L$-invariant matrix
\begin{eqnarray}\label{eta}
   \eta= \left(\begin{array}{cccc} \eta_{1}\otimes 1_{N_1 \times N_1} & 0 & \cdots &0 \\ 0 &\eta_{2}\otimes1_{N_2 \times N_2}&\cdots&0\\ 
\vdots &\vdots &\ddots&\vdots\\
0&0&\cdots&\eta_{M}\otimes1_{N_M \times N_M} \end{array}\right)_{N \times N} \,. 
\end{eqnarray}
In the case of ${\cal N}=4$ SYM some of the six real scalars of the theory may acquire vacuum expectation values (vevs). A maximally supersymmetric surface operator ${\cal O}_\Sigma$ induces for the complex scalar $\Phi={1\over \sqrt{2}}(\phi^5+i \phi^6)$ an $L$-invariant pole near the locus of the defect 
$\Sigma$ which reads
\begin{eqnarray}\label{Phi}
   \Phi= \frac{1}{ \sqrt{2}z}\left(\begin{array}{ccc} (\beta_{1}+i \gamma_1)\otimes 1_{N_1 \times N_1} & \cdots &0 \\
\vdots &\ddots&\vdots\\
0&\cdots&(\beta_{M}+i \gamma_M)\otimes1_{N_M \times N_M} \end{array}\right)_{N\times N}
\end{eqnarray}
%
where $z=r\, e^{i \psi}$ and $r$ is the normal distance from the defect.

 To summarise, a maximally supersymmetric surface operator  in ${\cal N}=4$ SYM
with Levi group $L=\prod_{l=1}^M U(N_l)$ is characterised by the $4M$
$L$-invariant parameters $(\alpha_l,\beta_l,\gamma_l,\eta_l)$.
Furthermore, the operators under consideration preserve the subgroup $PSU(1,1|2)\times PSU(1,1|2)\times U(1)\subset
PSU(2,2|4)$ of the 4-dimensional  superconformal group
of ${\cal N}=4$ SYM. Its bosonic subgroup is 
$SO(2,2)\times SO(2)\times SO(4)$, where $SO(2,2)$ is the conformal group in 2-dimensions, $SO(4)$ is the remaining R-symmetry group which rotates the 4 scalars $\phi_i'\, i=1,2,3,4$ and $SO(2)$ corresponds to a diagonal combination of
space-time and $R$-symmetries. \footnote{For Euclidean signature the spacetime symmetry instead of being $SO(2,2)$ is $SO(1,3)$.}
Finally, as shown in Refs.~\refcite{Gomis:2007fi,Gaiotto:2008sa} the surface operator defined by Eqs.~\eqref{AA}, \eqref{eta} and \eqref{Phi} is annihilated by 8 supersymmetric generators and another 8 superconformal generators respecting, thus, 16  of the 32 superconformal symmetries of the maximally supersymmetric theory in 4-dimensions.  

Based on the solution \eqref{AA}, \eqref{eta} and \eqref{Phi} one may compute several observables at weak coupling. One of the most significant observables in dCFTs is the 1-point function of chiral primary operators (CPOs) which acquires an non-zero value in the presence of the defect. From the symmetry group of the defect it is obvious that only the CPOs which are singlets under the $SO(4)$ R-symmetry group can have non-zero expectation values.
Two such unit normalised opertors are the following
\begin{equation}\label{CPOs}
    {\cal O}_{3,1}={8\pi^3\over \lambda^{3/2}}
\hbox{Tr}\Bigg(2\Phi^2\bar{\Phi}-\Phi\sum_{I=1}^4\phi^I\phi^I\Bigg)
\quad \text{and} \quad  
{\cal O}_{3,3}={32\pi^3\over\sqrt{6}\lambda^{3/2}}
\hbox{Tr}\left(\Phi^3\right) \, . 
\end{equation}
By inserting the classical solution \eqref{Phi} it is straightforward to calculate their 1-point function, see Ref. \refcite{Drukker:2008wr} 
\begin{eqnarray}\label{CPOs-rweak}
  &&  \langle {\cal O}_{3,1}\rangle ={1\over z|z|^2}
{ 8\pi^3\over \sqrt{2}\lambda^{3/2}}\sum_{l=1}^MN_l (\beta_l^2+\gamma^2_l)(\beta_l+i\gamma_l), \nonumber\\
&&\langle {\cal O}_{3,3}\rangle={1\over z^3}
{8\pi^3\over\sqrt{3}\lambda^{3/2}}\sum_{l=1}^MN_l (\beta_l+i\gamma_l)^3\,.
\end{eqnarray}

\subsection{The D3 probe brane}
The second description of a ${1 \over 2}$-BPS surface operator ${\cal O}_\Sigma$ is in terms of a probe D3 brane in $AdS_5\times S^5$ spacetime that ends on the boundary along the  surface of the defect $\Sigma$, first considered in Ref. \refcite{Constable:2002xt}.

The metric of the background written in Poincare coordinates reads
\begin{equation}\label{metric-Lor}
ds^2 = \frac{1}{z^2} \, \Big[-dx_0^2 + dx_1^2+ dr^2 + r^2 \, d\psi^2 + dz^2 \Big] +d\Omega_5^2
\end{equation}
with the length element on $S^5$ given by
\begin{equation}\label{metric}
d\Omega_5^2 = 
d\tilde\psi^2 + 
\sin^2 \tilde\psi \, \left(d{\tilde \beta}^2 + \sin^2 {\tilde \beta} \, d{\tilde \gamma}^2 \right) +
\cos^2 \tilde\psi \left(d\beta^2 + \sin^2 \beta \, d\gamma^2\right) \, . 
\end{equation}
In addition, we choose the RR 4-form potential to be given by
\begin{equation}\label{C4}
C_4 = \frac{r}{z^4} \, dx_0 \wedge dx_1 \wedge dr \wedge d\psi \, . 
\end{equation}
We choose the world-volume coordinates of the probr D3-brane to be $\zeta^\mu=(x_0,x_1,r,\tilde \gamma)$.
Then the solution of the equations of motion of the D3-brane for the remaining six coordinates takes the form
\begin{equation}\label{embedding}
\tilde\psi = {\pi\over 2} \, , \quad 
{\tilde \beta} = \frac{\pi}{2} ,\quad 
{\psi}+\tilde \gamma =\phi_0, \quad \beta=\frac{\pi}{2}, \quad \gamma=0 \quad \& \quad
z= \sigma \, r \, . 
\end{equation}

The D3-brane wraps an $S^1\subset S^5$ parametrized by the angle $\tilde \gamma$.
As a result, it is straightforward to show that the symmetry of the induced on the brane metric is 
$AdS_3\times S^1$ with the $AdS_3$ part corresponding to the conformal $SO(2,2)$ group mentioned in the previous section. 
The orientation of the codimension-2 D3-D3 probe-brane system is shown in Table \ref{Table:C2D3D5system}.  
\begin{table}[H]
\begin{center}\begin{tabular}{|c||c|c|c|c|c|c|c|c|c|c|}
\hline
& ${\color{red}x_0}$ & ${\color{red}x_1}$ & ${\color{red}r}$ & ${\color{red}\psi}$ & ${\color{red}z}$ & ${\color{blue}\tilde \psi}$ & ${\color{blue} {\tilde \beta} }$ & ${\color{blue} {\tilde{ \gamma}}}$ & ${\color{blue}\beta}$ & ${\color{blue}\gamma}$ \\ \hline
\text{D3} & $\bullet$ & $\bullet$ & $\bullet$ & $\bullet$ &&&&&& \\ \hline
\text{D3 probe} & $\bullet$ & $\bullet$ & $\bullet$ &  &   & & &$\bullet$ &  & \\ \hline
\end{tabular}
\caption{The D3-D3 brane intersection.
\label{Table:C2D3D5system}}\end{center}
\end{table}
\noindent
As mentioned above, the solution terminates on a 2-dimensional submanifold of the $AdS_5$ boundary, realising, thus, a codimension-2 dCFT. This is apparent since on the boundary which is at $z=0$, $r=0$ too. Consequently, the intersection of the D3 brane with $\partial AdS_5$ extends solely along $x_0$ and $x_1$, the coordinates parametrising the surface $\Sigma$ of the defect.

At this point, let us mention that the geometric data $\sigma$ and $\phi_0$ specifying the D3-brane solution are related to the parameters $\beta$ and $\gamma$ of the field theory side by, see Ref. \refcite{Drukker:2008wr}
\begin{equation}\label{rel}
 \beta_1 +i \gamma_1={\sqrt{\lambda}\over 2 \pi} {1\over \sigma}\, e^{i \phi_0}\, .   
\end{equation}
In addition, one may turn on  a Wilson line for the
gauge field $A$, as well as for the dual gauge field $\tilde{A}$ living on the
D3-brane worldvolume along the non-contractible $S^1$. Therefore,
the D3-brane solution ending on $\Sigma={\mathbb R}^2$ on the boundary
depends on two additional parameters $\alpha$ and $\eta$, which can be identified with those of the gauge theory
description of a surface operator ${\cal O}_\Sigma$ via
\begin{eqnarray}
\alpha_1=\oint {A\over 2\pi}, \qquad \eta_1=\oint {\tilde{A}\over 2\pi}\,.
\end{eqnarray}
So the single D3 probe brane is described by two stacks of branes, i.e. $M=2$. The first stack consists of a single brane characterised by the 4 parameters $(\alpha_1, \beta_1, \gamma_1, \eta_1)$, while the second stack consists of $N-1$ branes characterised by the $4(N-1)$ parameters which, however, have special values, namely 
$\beta_2+i \gamma_2=0$.
 
In the probe approximation one may calculate the vevs of CPOs by using the bulk-to-boundary propagator of the supergravity field dual to the CPO and by integrating its bulk end over the world volume of the D3 probe brane. The result for the CPOs of \eqref{CPOs} was obtained in Ref. \refcite{Drukker:2008wr} and reads
\begin{equation}\label{CPOs-rprobe}
     \langle {\cal O}_{3,1}\rangle={1\over\sqrt2}\,{\cosh^2u_0\sinh u_0\over r^3}\,e^{i(\phi_0-\psi')},\quad
 \langle {\cal O}_{3,3}\rangle
={1\over\sqrt3}\,{\sinh^3u_0\over r^3}\,e^{3i(\phi_0-\psi')}\,,
\end{equation}
where $z=r\,e^{i\psi'}$ and $\sinh u_0={1\over \sigma}$. By taking into account \eqref{rel}, as well as the fact that in the probe approximation the second stack of $N-1$ branes has $\beta_2+i \gamma_2=0$ it is easy to compare \eqref{CPOs-rprobe} and \eqref{CPOs-rweak}. The comparison shows that the weak and strong coupling results perfectly agree for $\langle {\cal O}_{3,3}\rangle$. For $\langle {\cal O}_{3,1}\rangle$ the result \eqref{CPOs-rprobe} written in field theory variables reads
\begin{equation}\label{comp-2}
    \langle {\cal O}_{3,1}\rangle ={1\over z|z|^2}
{ 8\pi^3\over \sqrt{2}\lambda^{3/2}} (\beta_1+i\gamma_1) 
\Bigg( (\beta_1^2+\gamma^2_1)+{\lambda \over (2 \pi)^2}\Bigg) \, .
\end{equation}
The first term in the parenthesis agrees precisely with the weak coupling result of \eqref{CPOs-rweak}. The structure of \eqref{comp-2} lead the authors of Ref. \refcite{Drukker:2008wr} to propose that the quantum corrections to the semiclassical
gauge theory result for the CPO ${\cal O}_{\Delta,k}$ stop  at order $\lambda^{(\Delta-|k|)/2}$.


\subsection{The bubbling supergravity solution}
A third description, perhaps the most complete, of ${1 \over 2}$-BPS surface operators ${\cal O}_\Sigma$ is in terms of the fully backreacted type IIB bubbling supergravity solution \refcite{Gomis:2007fi}, a solution which is asymptotically $AdS_5\times S^5$. The bubbling geometry describing the ${1 \over 2}$-BPS surface operators ${\cal O}_\Sigma$ is actually obtained from the bubbling geometries dual to the half-BPS local operators \refcite{Lin:2004nb} by performing a double “analytic” continuation. 

The strategy to derive the desired geometry is to seek for a solution that preserves the  same symmetries as the surface operator ${\cal O}_\Sigma$, namely $SO(2,2)\times SO(2)\times SO(4)$, see end of section \ref{GT}. The most general
10-dimensional metric invariant under these symmetries can be constructed by fibering $AdS_3\times S^1\times S^3$ realising the aforementioned group over a three manifold $X$ with the symmetries acting as isometries on the
fiber. The most general such ansatz is given by
\begin{eqnarray}\label{ansatz}
    ds^2&=&y\sqrt{2z+1\over 2z-1}ds^2_{AdS_3}+y\sqrt{2z-1\over 2z+1}d\Omega_3+{2y\over \sqrt{4z^2-1}}(d\chi+V)^2+\\
   &&+ {\sqrt{4z^2-1}\over 2y}(dy^2+dx_idx_i)\nonumber,
\end{eqnarray}
where $z(y,x_1,x_2)$ is a function defined on the space X which is parametrized by $x_1$, $x_2$ and $y$. The metric of $X$ is $ds^2_X = dy^2 + dx_idx_i$ and $y\geq 0$. Moreover, $V$ is an one-form in $X$ determined from $z$ by solving the
equation $dV={1\over y}*_X dz$. It is important to note that both the metric and the 5-form
are completely determined once the function $z(y,x_1,x_2)$ is specified.

A complete solution to the type IIB equations of motion is obtained by solving the following partial differential equation in the 3-dimensional $X$ space
\begin{eqnarray}\label{diff-eq}
\partial_i\partial_i z(x_1,x_2,y)+y\partial_y\left({\partial_y z(x_1,x_2,y)\over y}\right)=\sum_{l=1}^MQ_l\delta(y-y_l)\delta^{(2)}(\vec{x}-\vec{x}_l).
\end{eqnarray}
The solution is, thus, characterised by $M$ point particles of charge $Q_l,\, l=1,\cdots M$ each sitting at the point $(y_l,\vec{x}_l)$.

Equation \eqref{diff-eq} can now be solved with the boundary condition  $z(y=0,x_1,x_2)=1/2$ which imposes that the $S^3$ shrinks smoothly to a point at $y=0$.
The solution of \eqref{diff-eq}reads \refcite{}
\begin{equation}\label{sol-z}
 z(y,x_1,x_2)={1\over 2}+\sum_{l=1}^M z_l(y,x_1,x_2),
\end{equation}
with
\begin{equation}\label{sol-zl}
z_l(x_1,x_2,y)={(\vec{x}-\vec{x}_l)^2+y^2+y_l ^2
\over 2
\sqrt{((\vec{x}-\vec{x}_l)^2+y^2+y_l ^2 )^2-4y_l ^2y^2}}-{1\over 2}\,.
\end{equation}
As mentioned above, given the function $z$ the supergravity solution is fully determined. The value of the charge $Q_l$ is determined by the requirement that the $S^1$ parametrised by $\chi$ shrinks to zero size at the point $(y_l,\vec{x}_l)$. In order for this to happen in a smooth way the magnitude of the “charge” has to be fixed  so that $Q_l = 2\pi y_l$, see Ref. \refcite{Lin:2004nb}.

The solution \eqref{ansatz} has a rich topological structure. With each point particle at $(y_l,\vec{x}_l)$ one can associated a non-trivial 
5-sphere $S^5_l$ which can be constructed by fibering the $S^1\times S^3$ present in the geometry \eqref{ansatz} over a straight line between the
point $(0,\vec{x}_l)$ and the point $(y_l,\vec{x}_l)$ in $X$ (see Fig. 2 of Ref. \refcite{Drukker:2008wr}). The solution is such that at the point $(y_l,\vec{x}_l)$ the $S^1$ shrinks to zero size while at the point $(0,\vec{x}_l)$ it is the $S^3$ that shrinks to zero size. Furthermore, the boundary of the geometry \eqref{ansatz} which is at $y\rightarrow \infty$, $z\rightarrow {1\over 2}$ and the boundary has the geometry $AdS_3\times S^1$ and possesses a non-contractible circle $S^1$. Now one may fiber 
the $S^1$ parametrized by $\chi$  over a straight line connecting the point $(y_l,\vec{x}_l)$ in X – where
the $S^1$ shrinks to zero size – to the point in X corresponding to the boundary of $AdS_5\times S^5$
– given by $(\infty,\vec{x}_l)$ – to obtain a surface which is topologically equivalent to a disc $D_l$. So although the $S^1$ is not contractible on the boundary it is when considering the bulk.

 Thus, in order to fully determine the bubbling solution of   Type IIB supergravity one must, in addition to the metric and the 5-form, specify the integral of the NS-NS and R-R 2-forms through the disks $D_l$
\begin{eqnarray}\label{a-eta}
   \alpha_l=-\int_{D_l} {B_{NS}\over 2\pi},\qquad \qquad \eta_l=\int_{D_l} {B_{R}\over 2\pi},\qquad l=1,\ldots,M \, .
\end{eqnarray}
We conclude that the bubbling supergravity solution dual to the surface operators ${\cal O}_\Sigma$ depends on the same number of parameters $4M$, as many as those needed for the description of ${\cal O}_\Sigma$ in gauge theory. These parameters are $(\alpha_l,\beta_l,\gamma_l,\eta_l), \, l=1,\cdots M$ and they are related to the gauge theory ones by \eqref{a-eta} and 
\begin{equation}\label{rel-2}
 \beta_l+i \gamma_l={\sqrt{\lambda}\over 2 \pi} (x_{1,l}+i x_{2,l}). 
\end{equation}

 Finally, by using the bubbling geometry and \eqref{rel-2} one may write the results for the 1-point function of the CPOs of \eqref{CPOs} as 
\begin{eqnarray}\label{CPOs-rsugra}
&&\langle {\cal O}_{3,1}\rangle={1\over z|z|^2}
{8\pi^3\over\sqrt{2}  \lambda^{3/2}}
\left(
\sum_{l=1}^M N_l\left({(\beta_l^2+\gamma_l^2)}
+{\lambda\over 4\pi^2} {N-2N_l\over 2N}\right)(\beta_l+ i\gamma_l)\right)
\nonumber\\
&&\langle {\cal O}_{3,3}\rangle={1\over z^3}
{8\pi^3\over \sqrt{3}\lambda^{3/2}}\sum_{l=1}^M
N_l(\beta_l+ i\gamma_l)^3.    
\end{eqnarray}
 In order to compare the supergravity result with the result in the the probe brane approximation \eqref{CPOs-rprobe}, we should set in 
 \eqref{CPOs-rsugra} $N_1=1$ and $N_2=N-1$ and 
take the $N\rightarrow \infty$ limit. Then the results are in precise agreement. 

Other observables than can and have been calculated for this class of operators in any of the three descriptions is the 1-point function of the stress-energy tensor, the 1-point function of R-currents, the expectation values of Wilson  Ref.~\refcite{Drukker:2008wr}, as well as the anomaly coefficients associated with the presence of the defect.

\section{Holographic interpolations of  dCFTs}
\label{Holo_interpolations}

In this section, we review material from Refs.~\refcite{Georgiou:2025mgg} and \refcite{Georgiou:2025wbg}. Reversing the chronological order, we begin with Ref. \refcite{Georgiou:2025wbg}, which introduces a novel class of holographic dualities involving generically non-supersymmetric defect conformal field theories (dCFTs) and their gravitational counterparts. After establishing the duality, we focus on a specific example within this class to calculate the one-point functions of the energy-momentum tensor and the chiral primary operators (CPOs) at both strong and weak coupling. We then demonstrate that, in an appropriate limit, a compelling agreement emerges between the two regimes.


\subsection{The D5 probe brane}

\begin{figure}[t]
\centering
\includegraphics[width=6cm,height=5cm]{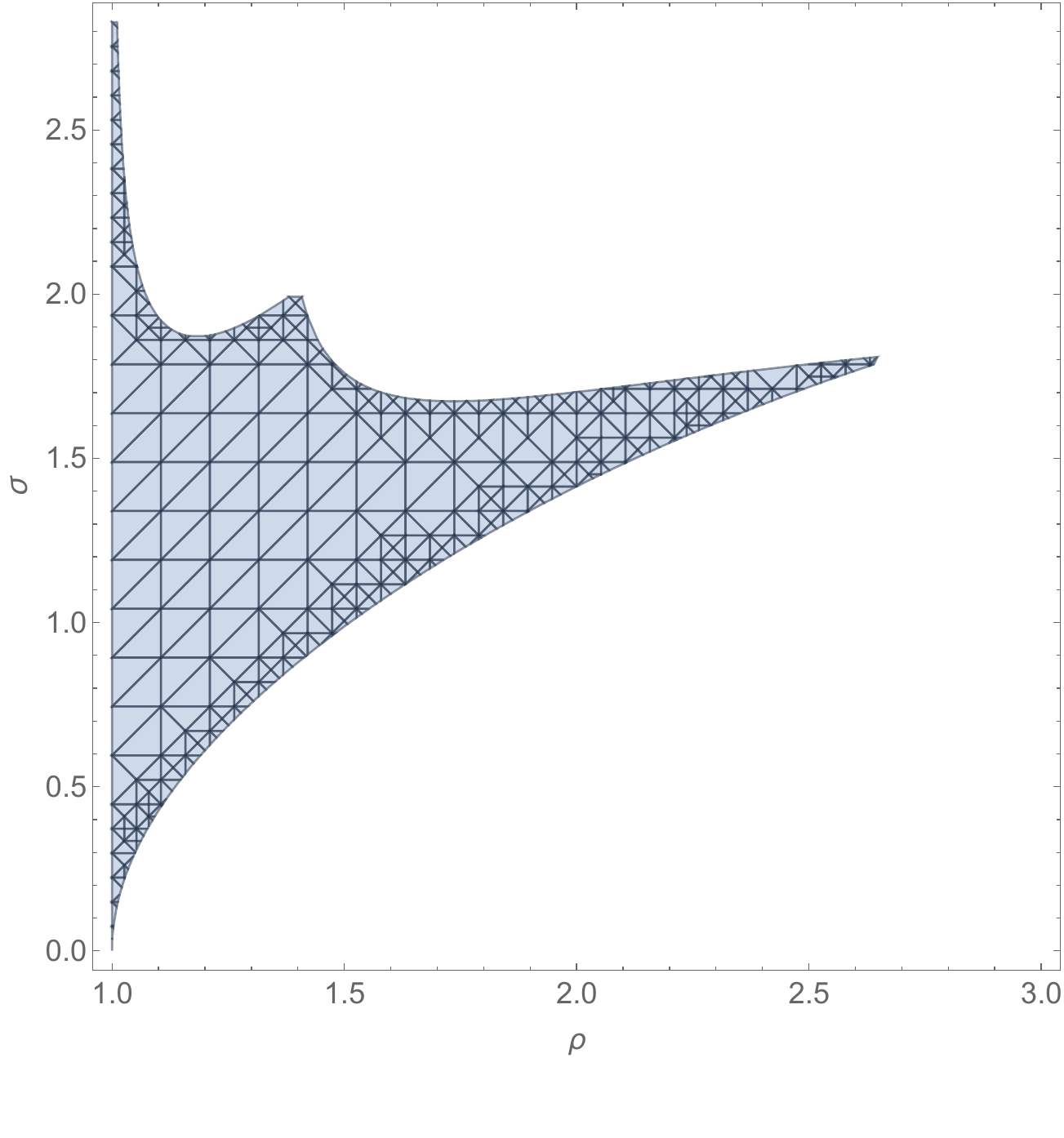}
\includegraphics[width=6cm,height=5cm]{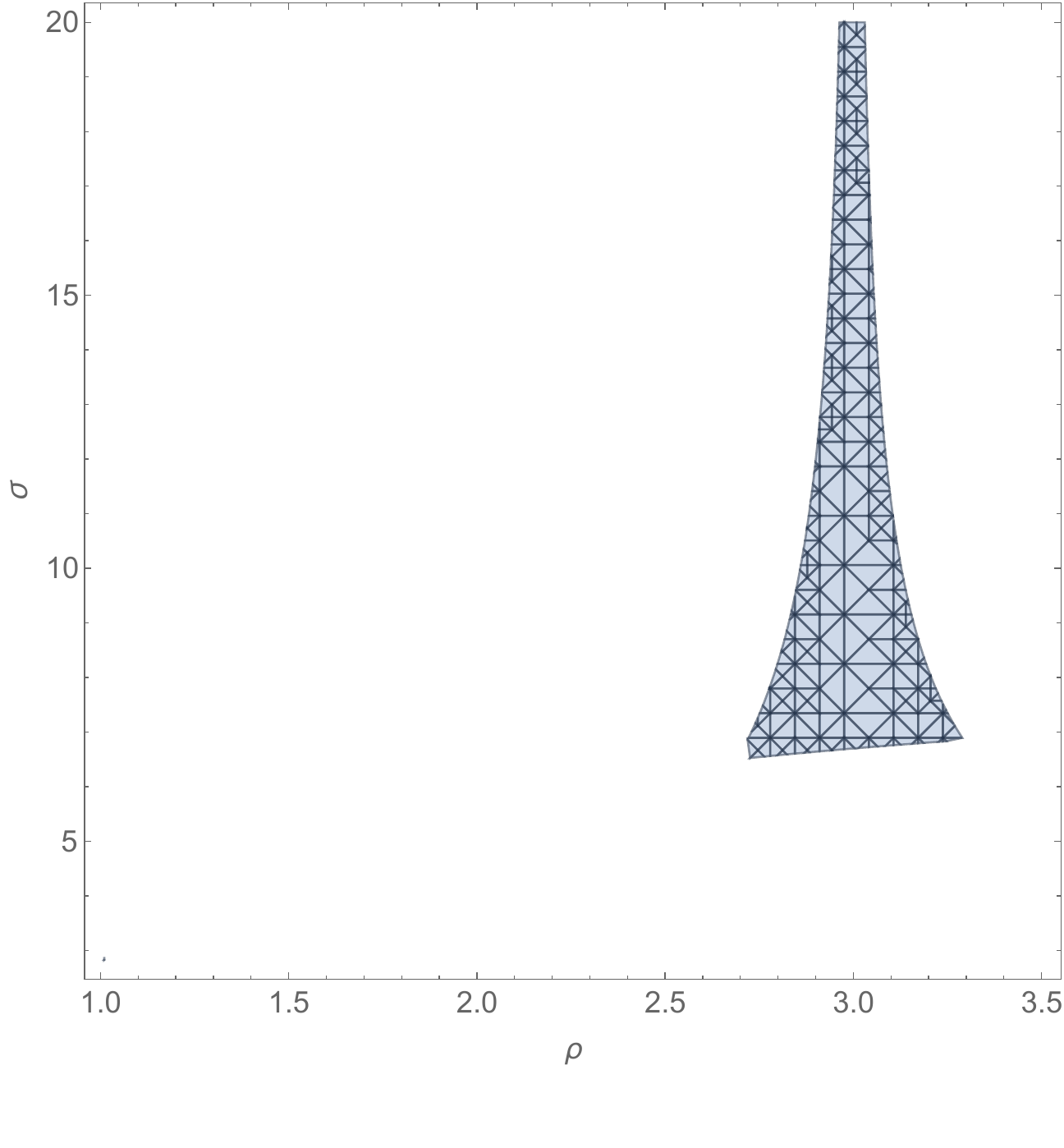}
\caption{The regions of the $(\sigma,\rho)$
parametric space where the proposed solution remains valid. 
The left panel covers the range  $\sigma \in [0,2 \sqrt{2}]$ with 
$1 < \rho \le 3$, while the right panel illustrates the domain 
$\sigma \geq 2 \sqrt{2}$ and $1 < \rho \le 3.5$.}
 \label{figg-1}
 \end{figure}

In this section, we focus on the gravity side and present a novel D5-brane solution embedded in the $AdS_5\times S^5$ geometry. 
Our solution depends on two independent parameters and terminates on a two-dimensional submanifold of the $AdS_5$ boundary, thereby realizing a codimension-2 dCFT. 
The D5-brane wraps an $S^2$ within the internal $S^5$ space and extends along an $S^1\subset S^5$ parametrized by the angle $\tilde \gamma$.
Consequently, the symmetry of the induced metric on the brane is 
$AdS_3\times S^1\times S^2$. 
The orientation of the codimension-2 D3-D5 probe-brane system is detailed in Table \ref{Table:C2D3D5systemv2}, with the D5-brane worldvolume coordinates denoted as $\zeta^\mu = (x_0,x_1,r,\tilde \gamma, \beta, \gamma)$.  
\begin{table}[H]
\begin{center}\begin{tabular}{|c||c|c|c|c|c|c|c|c|c|c|}
\hline
& ${\color{red}x_0}$ & ${\color{red}x_1}$ & ${\color{red}r}$ & ${\color{red}\psi}$ & ${\color{red}z}$ & ${\color{blue}\tilde \psi}$ & ${\color{blue} {\tilde \beta} }$ & ${\color{blue} {\tilde{ \gamma}}}$ & ${\color{blue}\beta}$ & ${\color{blue}\gamma}$ \\ \hline
\text{D3} & $\bullet$ & $\bullet$ & $\bullet$ & $\bullet$ &&&&&& \\ \hline
\text{D5 probe} & $\bullet$ & $\bullet$ & $\bullet$ &  &   & & &$\bullet$ & $\bullet$ & $\bullet$ \\ \hline
\end{tabular}
\caption{The D3-D5 intersection.
\label{Table:C2D3D5systemv2}}\end{center}
\end{table}
\noindent
The embedding ansatz for the D5-brane is
\begin{equation}\label{embeddingv2}
\tilde\psi = \tilde\psi_0 \, , \quad 
{\tilde \beta} = \frac{\pi}{2} ,\quad 
{\psi} =\rho \,\tilde \gamma +\phi_0\quad \& \quad
z= \sigma \, r \quad {\rm with } \quad \sigma>0
\end{equation}
and the solution is supported by a worldvolume 2-form flux
\begin{equation}\label{A}
F = - \frac{\kappa}{2 \, \pi \, \alpha'} \, \sin \beta \, 
d\beta \wedge d\gamma 
\quad \text{with} \quad k=-\int_{S^2} \frac{F}{2 \, \pi} \quad \Rightarrow \quad 
k=\frac{\kappa}{\pi \alpha'}=\frac{\kappa \sqrt{\lambda}}{\pi}
\end{equation} 
where $k \in {\mathbb N}^*$.
Given the ansatz in Eq. \eqref{embeddingv2}, the equations of motion derived 
from the D5-brane action, comprised by the sum of the DBI and WZ terms, determine the constants 
$\kappa$ and $\tilde \psi_0$ in terms of $\sigma$ and $\rho$. 
Here, $\sigma$ determines the inclination angle of the D5-brane 
with respect to the $AdS$ boundary, while $\rho$ characterizes 
the winding of the brane around one of the internal angles. 
The condition $0\leq \cos{\psi_0}\leq 1$ implies that $\rho\ge 1$.
By examining the stability of the configuration-specifically by 
requiring that the masses of all fluctuations transverse to the D5-brane 
satisfy the Breitenlohner–Freedman bound-we determine the stable regions 
in the $(\rho,\sigma)$ parameter space, as depicted in Fig.~\ref{figg-1}.

Within the D5-brane framework, the radii of the $S^2$ and $S^1$ 
subspaces are governed by $\cos{\tilde \psi_0}$ and $\sin{\tilde \psi_0}$,
respectively. The solution admits two distinct limiting configurations. 
In the first case, as $\cos{\tilde \psi_0} \rightarrow 1$, the $S^2$ 
reaches its maximal size while the $S^1$ collapses to a point. In this regime, the induced metric on the brane is
\begin{equation}\label{induced-metric-limit1}
ds^2_{ind}=\frac{1}{r ^2 \sigma^2} 
\Bigg[-d{\vec x}_{0,1}^2+(1+\sigma^2)dr^2 \Bigg] 
+\frac{d\psi^2}{\sigma^2}+d\Omega^2_{\beta,\gamma} \quad {\rm with} \quad  \kappa=\frac{4+\sigma^2}{\sigma \sqrt{8-\sigma^2}}
\end{equation}
matching the results previously established for the D5-brane in 
Ref. \refcite{Georgiou:2025mgg}. 
Conversely, the second endpoint is approached when 
$\cos{\tilde \psi}_0\rightarrow 0$.
Here, the two-sphere vanishes, causing the D5-brane to effectively reduce to a D3-like configuration with an induced metric of the form
\begin{equation}\label{induced-metric-limit2}
ds^2_{ind}=\frac{1}{r^2 \sigma^2} 
\Bigg[-d{\vec x}_{0,1}^2+(1+\sigma^2)dr^2\Bigg]+ 
\left(1+\frac{1}{\sigma^2}\right) d\tilde{\gamma}^2 
\quad \text{with} \quad \kappa = 0 \, . 
\end{equation}
This configuration precisely recovers the supersymmetric D3 probe-brane, which serves as the holographic dual for Gukov-Witten surface 
operators in Refs. \refcite{Drukker:2008wr,Gukov:2006jk}. 
It is interesting to observe that although the D5-brane solution 
breaks all supersymmetries of 
$AdS_5\times S^5$, the $\cos\tilde\psi_0 \rightarrow 0$
limit restores supersymmetry, resulting in a 1/2 BPS object 
that is consistent with the emergent D3-brane geometry.

The existence of boundaries in the D5 probe brane, 
caused by the non-integer value of the parameter $\rho$ in Eq.  
\eqref{embedding}, leads to a violation of the gauge 
invariance of the action. To resolve this inconsistency, 
following Ref. \refcite{Georgiou:2025wbg},
it is proposed that placing two D7-branes at the points where the D5 brane
terminates could lead to the cancellation of the anomalous terms through an anomaly inflow mechanism from the D7 branes. 
Notably, the effect of the D7 branes is minimal, as 
they do not introduce expectation values for the ${\cal N}=4$ bulk scalars, 
but only for the scalar fields of the hypermultiplets that live on the defect.


\subsection{Conformal field theory dual}

In this section, we discuss the details of the defect CFTs that are dual to the D5-brane solutions. These CFTs are realized as classical solutions to the ${\cal N}=4$ SYM equations of motion, describing the two-dimensional interface generated by the D5-brane.

The Lagrangian density of ${\cal N}=4$ SYM is given by
\begin{equation} \label{LagrangianSYM}
{\cal L}_{{\cal N} = 4} = \frac{2}{g_{\text{\scalebox{.8}{YM}}}^2} \text{tr}\bigg\{-\frac{1}{4} F_{\mu\nu}^2- \frac{1}{2} \left(D_{\mu}\varphi_i\right)^2 + \frac{i}{2}\,\bar{\psi}\slashed{D}\,\psi +\frac{1}{2}\,\bar{\psi}\,\Gamma^{i+3}\, \left[\varphi_i, \psi \right]+ \frac{1}{4}\left[\varphi_i,\varphi_j\right]^2\bigg\} 
\end{equation}
where $\bar{\psi}_{\alpha} \equiv \psi_{\alpha}^{\dagger} \Gamma^0$ and
$\slashed{D} \equiv \Gamma^{\mu}D_{\mu}$. 
The gauge field strength and covariant derivatives are defined as
\begin{IEEEeqnarray}{l}
F_{\mu\nu} \equiv \partial_{\mu}A_{\nu} - \partial_{\nu}A_{\mu} - i \left[A_{\mu},A_{\nu}\right] \quad {\rm and} \quad 
D_{\mu}f \equiv \partial_{\mu}f - i \left[A_{\mu},f\right] \, .  \label{CovariantDerivatives}
\end{IEEEeqnarray}
In this context, $\Gamma^{M}$ and $\psi$ denote the 10-dimensional 
gamma matrices and fermion fields, respectively, while $\varphi_i$
represent the six real scalar fields of $\mathcal{N}=4$ SYM. 
All fields transform in the adjoint representation of $\mathfrak{u}(N)$.
The equations of motion for the bosonic fields are given by
\begin{IEEEeqnarray}{c}\label{eoms}
D^{\mu}F_{\mu\nu} = i\left[D_{\nu}\varphi_i,\varphi_i\right]
\quad \text{and} \quad 
D^{\mu}D_{\mu}\varphi_i = 
\Big[\varphi_j,\left[\varphi_j,\varphi_i\right]\Big].
\end{IEEEeqnarray}


A novel class of non-supersymmetric defect operators in ${\cal N}=4$
SYM is described in the bulk through the probe D5-brane solutions of the
previous subsection. These operators are supported on the 
$\Sigma= \mathbb{R}^{(1,1)}$ surface, which is parameterized by the 
coordinates $(x_0,x_1)$. 
Such non-local surface operators, denoted as ${\cal O}_\Sigma$, 
induce a codimension-2 singularity in the classical fields of 
${\cal N}=4$ SYM. The gauge field singularity associated with these surface operators manifests as a non-Abelian vortex configuration, which in the present context is given by
\begin{equation}\label{A-sing}
A= {\cal A}\, d\psi 
\quad \text{with} \quad 
{\cal A} =\left(\begin{array}{cc} a_1 \otimes I_{(N_1+k) \times (N_1+k)} &\,\,\,\, 0_{(N_1+k) \times \left(N -N_1-k \right)} \\ 0_{\left(N - N_1-k\right)\times (N_1+k)} &\,\,\,\,   0_{\left(N -N_1- k\right)\times \left(N-N_1 - k\right)} \end{array}\right)_{N\times N}\,
\end{equation}
where $N_1$ denotes the number of D5 branes and $k$ represents the 
integer flux through the $S^2_{(\beta,\gamma)}$, of the D5 solution.
It should be noted that, within the probe approximation, the lower-right block of $\cal A$ is taken to be zero.

The gauge singularity in Eq. \eqref{A-sing} breaks the $G=U(N)$ 
gauge group down to $U(N_1+k)\times U(N-N_1-k)$. 
In this context, $\psi$ represents the polar angle in the plane transverse to the defect surface  $\Sigma$, while the parameter  $a_1$ characterizes the surface operator. 
Away from the operator's location, the gauge field  in \eqref{A-sing} satisfies the corresponding equation of motion (the first equation in \eqref{eoms}), provided that the commutator on the right-hand side vanishes. As will be shown, this condition is indeed satisfied. The resulting field strength is then given by
\begin{equation}
F=dA=2 \pi {\cal A} \, \delta_\Sigma  \,\,\,  {\rm where}  \,\,\,  \delta_\Sigma=d(d \psi)
\end{equation}
and, as previously noted, it satisfies the gauge field equation of motion. In this expression, $\delta_\Sigma=d(d\psi)$ represents a two-form delta function supported on the surface of the operator.

To identify the scalar fields that acquire vacuum expectation values (vevs), we first observe that, within the setup that is described in 
Table \ref{Table:C2D3D5system}, the six coordinates of the $S^5$, $X_i,\,\,i=1,\cdots 6$ satisfy $X_1^2+X_2^2+X_3^2=\sin^2{\tilde \psi}$ and $X_4^2+X_5^2+X_6^2=\cos^2{\tilde \psi}$. 

The symmetries of the system motivate the following ansatz for the bulk scalars
\begin{align}\label{sol-2}
\varphi_{i+3}^{\text{cl}}\left(r'\right) &= \frac{1}{\sqrt{2}\,r'}   \cdot \left[\begin{array}{cc} \left(t_i\right)_{k\times k} & 0_{k\times \left(N - k\right)} \\ 0_{\left(N - k\right)\times k} & 0_{\left(N - k\right)\times \left(N - k\right)} \end{array}\right], \quad i=1,2,3 
\\[5pt]
 \varphi_{1}^{\text{cl}}\left(r'\right) &= 0 
 \quad \& \quad \varphi_{2}^{\text{cl}}(r',\psi) +i \varphi_{3}^{\text{cl}} (r',\psi)=   \sin{\tilde \psi}_0\frac{1}{z}\,C \, . \label{sol-2a}
\end{align}
In this expression, $r'=\sqrt{x'^2_2+x'^2_3}$
represents the radial distance from the defect, while the complex coordinate is defined as $z=r' e^{i\psi}$. 
Furthermore, $C$ denotes an $N\times N$ matrix given by
\begin{equation} \label{C-matr}
C=\left(\begin{array}{cc} (\beta_1+i \gamma_1) \otimes I_{(N_1+k)\times (N_1+k)} &\,\,\,\, 0_{(N_1+k)\times \left(N-N_1 - k\right)} \\ 0_{\left(N-N_1- k\right)\times (N_1+k)} &\,\,\,\,  0_{\left(N-N_1 - k\right)\times \left(N-N_1 - k\right)} \end{array}\right)_{N\times N}\, .
\end{equation}
The matrices $t_i$ constitute a $k$-dimensional irreducible representation of $\mathfrak{su}\left(2\right)$
\begin{equation}\label{tmatr-comm}
\left[t_i, t_j\right] = i\,\epsilon_{ijl}\,t_l, \quad i,j,l = 1,2,3
\quad \text{with} \quad
t_i=\frac{1}{2}\sum_{j=1}^3\left[t_j,\left[t_j,t_i\right]\right] \, . 
\end{equation} 
The expressions in Eqs. \eqref{sol-2} and \eqref{sol-2a}, which determine the classical profiles for the six scalar fields, must satisfy the 
${\cal N}=4$ SYM equations of motion presented in Eq. \eqref{eoms}. Notably, the gauge field $A$ defined in Eq. \eqref{A-sing} commutes with itself and with the scalar vacuum expectation values (vevs) in Eqs. \eqref{sol-2} and \eqref{sol-2a}. Consequently, the covariant derivatives reduce to standard partial derivatives, i.e., $D_\mu=\partial_\mu$, on the solution.

Given these considerations, the equations of motion in Eq. \eqref{eoms} simplify to:\footnote{The equations of motion hold everywhere except 
from the location of the surface operator.}
\begin{IEEEeqnarray}{c}\label{eoms1}
\left[\partial_{\nu}\varphi_i,\varphi_i\right]=0, \qquad \partial^{\mu}\partial_{\mu}\varphi_i = \Big[\varphi_j,\left[\varphi_j,\varphi_i\right]\Big].
\end{IEEEeqnarray}
Reflecting the symmetry of our D5-brane solution, we assume the scalar fields depend exclusively on the radial distance $r'$ and the polar angle 
$\psi$, which parameterize the plane transverse to the surface 
$\Sigma$. Under the assumption $\varphi_i=\varphi_i(r',\psi)$, 
the second expression in Eq. \eqref{eoms1} further reduces to
\begin{equation}\label{radial}
\frac{\partial^2\varphi_i}{\partial r'^2}+\frac{1}{r'}\frac{\partial \varphi_i}{\partial r'}+\frac{1}{r'^2}\frac{\partial^2 \varphi_i}{\partial \psi^2}= 
\Big[\varphi_j,\left[\varphi_j,\varphi_i\right]\Big].
\end{equation}
It is evident that the ansatz in Eqs. \eqref{sol-2} and \eqref{sol-2a} satisfies the first condition in Eq. \eqref{eoms1}. 
Regarding Eq. \eqref{radial}, we observe that the scalar fields involving the matrix $C$ solve the corresponding homogeneous equation. 
This is sufficient, as all commutators involving 
$ \varphi_{1}^{\text{cl}} $, $ \varphi_{2}^{\text{cl}} $ and $ \varphi_{3}^{\text{cl}} $ vanish. The remaining non-zero commutators involve only $ \varphi_{4}^{\text{cl}} $, $ \varphi_{5}^{\text{cl}}\ $ and $ \varphi_{6}^{\text{cl}} $.
By applying Eq. \eqref{tmatr-comm}, it can be demonstrated that Eq. \eqref{radial} is also satisfied for these scalars. In both cases, we have utilized the property $[\varphi_i,\varphi_{i+3}]=0$ for $i=1,2,3$.

In summary, the scalar field equations of motion are satisfied by Eqs. \eqref{sol-2} and \eqref{sol-2a}. Two points are worth noting. 
First, these solutions—which provide the dual description of the D5 probe brane—interpolate between two regimes: the dCFT dual to the half-BPS D3-D3 system of Ref. \refcite{Gukov:2006jk} at $\tilde\psi_0=\pi/2$, 
and the non-supersymmetric dCFT of Ref. \refcite{Georgiou:2025mgg} 
at $\tilde\psi_0=0$.\footnote{To achieve exact correspondence, the matrices in Ref. \refcite{Georgiou:2025mgg} should be rescaled as $t_i\rightarrow t_i/\sqrt{2}$.}
Specifically, in the limit $\tilde\psi_0=\pi/2$, where 
$k\sim \kappa\rightarrow 0$, the expression for the matrix 
in Eq. \eqref{sol-2} becomes meaningless; this implies that 
$ \varphi_{4}^{\text{cl}} $, $ \varphi_{5}^{\text{cl}} $ 
and $ \varphi_{6}^{\text{cl}}$ should vanish. 
Conversely, at the $\tilde\psi_0=0$ endpoint, we find that 
$\varphi_{1}^{\text{cl}}=\varphi_{2}^{\text{cl}}=\varphi_{3}^{\text{cl}}=0$,
while the remaining three scalars coincide with the non-vanishing scalar fields of the dCFT in Ref. \refcite{Georgiou:2025mgg}. 

To verify the proposed duality between the holographic description involving the probe D5-brane and the field theory solution presented above, we compute the one-point functions of the energy-momentum tensor and the chiral primary operators in the subsequent subsections. For simplicity, these computations are performed in the limit $\tilde\psi_0=0$, where the D5-brane serves as the gravity dual to the dCFT of Ref. \refcite{Georgiou:2025mgg}.


\subsection{One-point function of the stress-energy tensor}

In the case of a codimension-two defect CFT, the residual conformal 
symmetry constrains the spacetime structure of the energy-momentum 
tensor's one-point function, determining it up to a function 
${\mathbf h}={\mathbf h}(g_{YM},N)$. 
For an operator localized at the boundary point 
$x^m=\{ x_0,x_1,x_2,x_3 \}=\{ 0,0,0,r' \}$, 
the only non-vanishing components of the vacuum expectation 
value (vev) of $T_{\mu\nu}$ are 
\begin{equation} \label{nonzeroT}
\langle T_{00}\rangle=\langle T_{11}\rangle=\langle T_{33}\rangle= \frac{{\mathbf h}}{r'^4}, \qquad \langle T_{22}\rangle=-3 
\frac{{\mathbf h}}{r'^4} \quad \text{with} \quad 
\langle T_{\,\,\, m}^m\rangle=0 \, . 
\end{equation}
The function $\mathbf h$ encodes the specific features of the 
defect CFT under consideration. 
In the remainder of this section, we determine its form at strong coupling by employing the holographic correspondence, and at weak coupling within the framework of ${\cal N}=4$ SYM theory.

\subsubsection{Holographic computation}

In this subsection, we perform the holographic computation of the one-point function for the energy-momentum tensor in the presence of a 
codimension-two defect CFT. 
This observable is of particular interest, as it vanishes both in the absence of a defect and in theories involving defects of codimension one.

\begin{figure}[t]
\centering
\includegraphics[width=6cm,height=5cm]{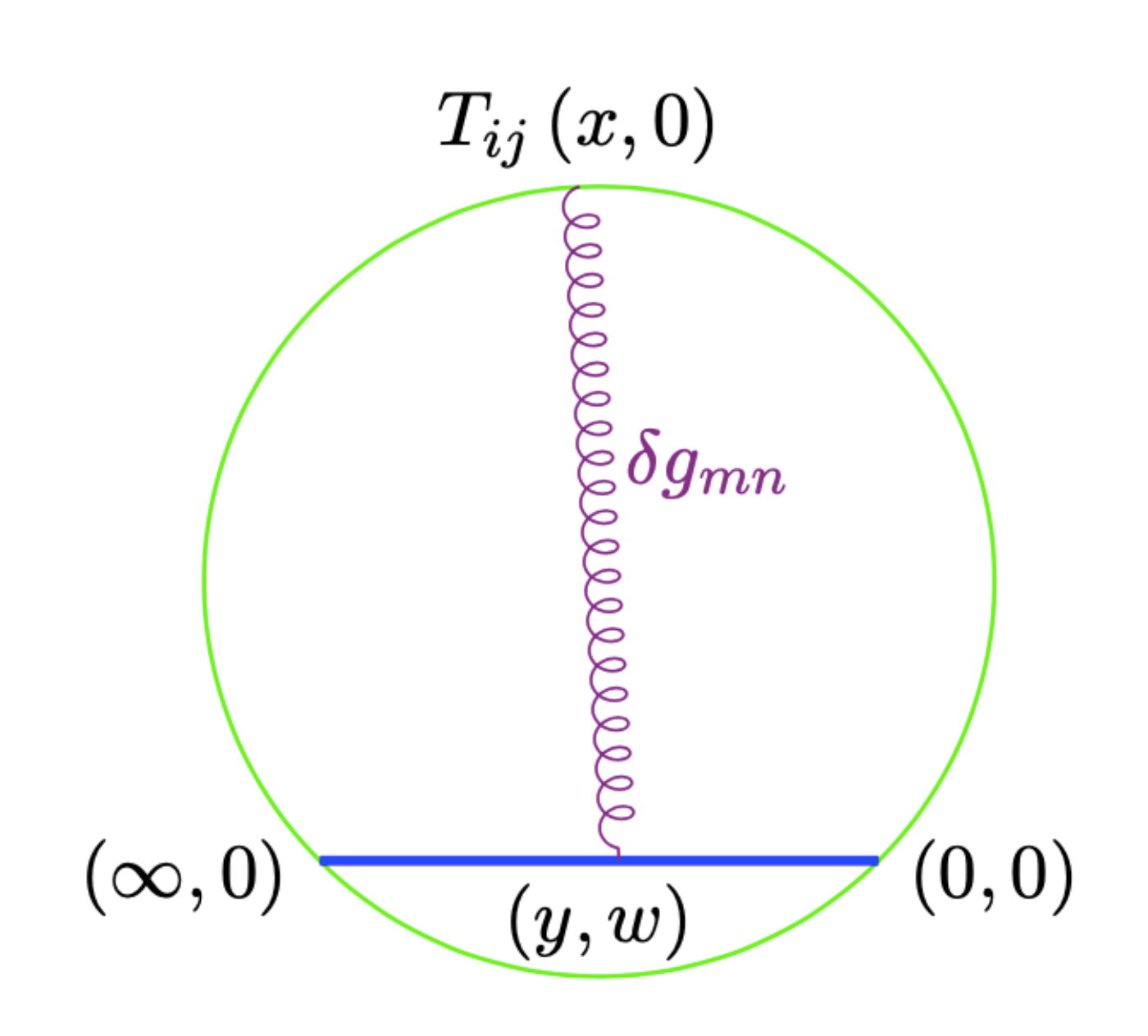}
\includegraphics[width=6cm,height=5cm]{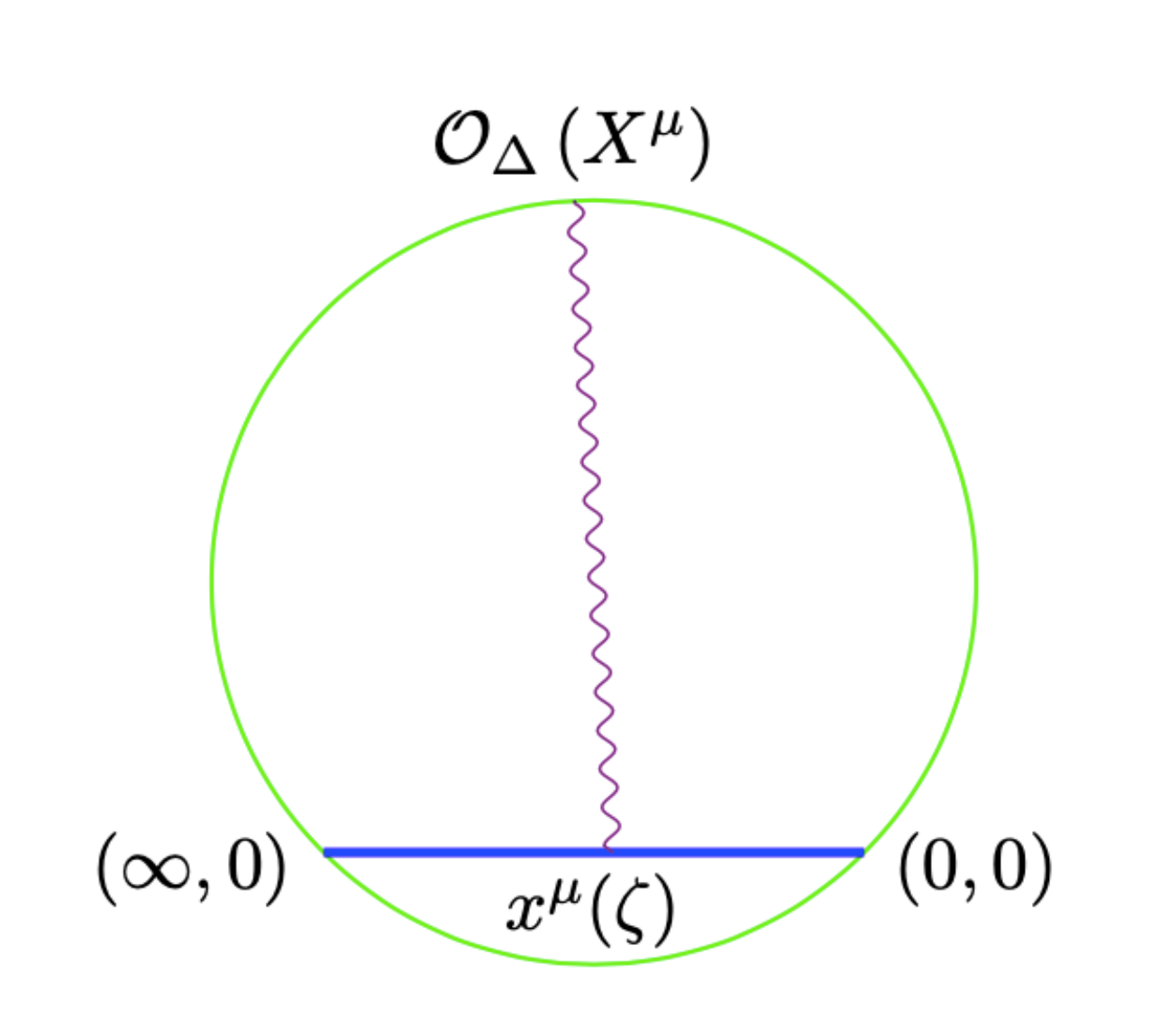}
\caption{Left panel: One-point function of the energy-momentum tensor in a defect CFT. The green circle represents the $AdS_5$ boundary. The wavy line illustrate the fluctuations of the metric. 
Right panel: One-point function of a CPO in a defect CFT, where the operator is localized at the boundary point $X^\mu=(X_0,X_1,r',\psi',0)$. Here, $x^\mu(\zeta)$ denotes the spacetime coordinates of an arbitrary point on the D5-brane worldvolume, parametrized by $\zeta=(x_0,x_1,r,\psi,\beta,\gamma)$. 
The wavy line illustrate the fluctuations of the metric and the 4-form potential.}
\label{figg-2}
\end{figure}

The IIB supergravity field dual to the stress tensor is the fluctuation 
of the AdS metric: $g_{mn} = \hat{g}_{mn} + \delta g_{mn}$,
where $\hat{g}_{mn}$ denotes the background metric of AdS$_{d+1}$ 
containing the probe Dp-brane. 
To evaluate the one-point function at strong coupling, we follow the prescription of Ref. \refcite{Georgiou:2023yak}. 
In the presence of a ``heavy" object—specifically the Dp-brane—the one-point function of the stress-energy tensor is given by
\begin{IEEEeqnarray}{l}
\big\langle T_{mn}\left(x\right)\big\rangle_{\text{brane}} = \lim_{z\rightarrow 0}\Big\langle\delta g_{mn}\left(x,z\right)\cdot \frac{1}{Z_{\text{brane}}}\int D \mathbb{Y}\,e^{-S_{\text{brane}}\left[\mathbb{Y}\right]}\Big\rangle_{\text{bulk}} \, . \label{OnePointFunctionsDpBrane1}
\end{IEEEeqnarray}
In the strong coupling limit, the path integral in Eq. \eqref{OnePointFunctionsDpBrane1} is dominated by a saddle point associated with the classical solutions $\mathbb{Y}_{\text{cl}}$, 
which describe the Dp-brane embedding in the $AdS_5\times S^5$ 
spacetime. Expanding the brane action around this classical configuration cancels the partition function in the denominator, 
and the one-point function becomes
\begin{IEEEeqnarray}{l}
\big\langle T_{mn}\left(x\right)\big\rangle_{\text{brane}} = -\lim_{z\rightarrow 0}\Bigg\langle\delta g_{mn}\left(x,z\right)\cdot\left(\left.\frac{\partial S_{\text{brane}}\left[\mathbb{Y}_{\text{cl}}\right]}{\partial\delta g_{kl}}\right|_{\delta g_{kl} = 0}\cdot\delta g_{kl}\left(y,w\right)\right) \Bigg\rangle_{\text{bulk}}, \label{OnePointFunctionsDpBrane2}
\end{IEEEeqnarray}
where $\left(x,0\right)$ is a point on the AdS boundary representing the operator's insertion, while $\left(y,w\right)$
denotes a point on the Dp-brane (see Fig. \ref{figg-2}). 
The general variation of the probe action with respect to 
the metric $g_{mn}$ is given by
\begin{IEEEeqnarray}{l}
\frac{\partial S_{\text{brane}}\left[\mathbb{Y} \right]}{\partial\delta g_{mn}} = \frac{T_p}{2g_s}\int \left[d^{p+1}\zeta \sqrt{h} \, h^{ab} \partial_a \mathbb{Y}^{m} \partial_b \mathbb{Y}^{n}\right], \label{DBIvariation}
\end{IEEEeqnarray}
where $h_{ab}$ is the induced metric.
Substituting these results into Eq. \eqref{OnePointFunctionsDpBrane2} and utilizing the bulk-to-boundary graviton propagator, we obtain a six-dimensional integral. Upon integration, the resulting stress tensor one-point function matches the form of Eq. \eqref{nonzeroT} required by conformal invariance, with the function ${\mathbf h}$ determined as
\begin{equation}\label{hT}
{\mathbf h}^{(strong)}=-\frac{ \lambda ^{3/2} \left(2 \sigma ^2+1\right)}{6 \pi ^3 \sigma ^3 \sqrt{8-\sigma ^2} g_{YM}^2}=-\frac{ \lambda ^{1/2} N \left(2 \sigma ^2+1\right)}{6 \pi ^3 \sigma ^3 \sqrt{8-\sigma ^2} }
\end{equation}
where $N$ is the number of colors. 

\subsubsection{Weak coupling computation \& agreement with holography}

We now proceed to the weak coupling analysis of the energy-momentum tensor. For this purpose, we employ the improved energy-momentum tensor derived from the Lagrangian in Eq. \eqref{LagrangianSYM}. The components relevant to our study are those involving only scalar fields, as provided in Refs. \refcite{Callan:1970ze,Sohnius:1981sn}
\begin{equation} \label{StressTensorScalar}
T_{\mu\nu}  =  \frac{2}{g_{\text{\scalebox{.8}{YM}}}^2} \cdot
\text{tr} \Bigg[
\frac{2}{3}
\left(\partial_{\mu}\varphi_i\right)
\left(\partial_{\nu}\varphi_i\right) 
- \frac{1}{3}\,\varphi_i\left(\partial_{\mu}\partial_{\nu}\varphi_i\right) 
- \frac{\eta_{\mu\nu}}{6} \, 
\Big[\left(\partial \varphi_i\right)^2 + \frac{1}{2}\left[\varphi_i,\varphi_j\right]^2\Big] \Bigg]
\end{equation}
where $i=1,2,\dots,6$.
The one-point function of the energy-momentum tensor is subsequently obtained by substituting the classical solution of Eq. \eqref{sol-2} into Eq. \eqref{StressTensorScalar}. 
It should be noted that in the present D5-brane setup, the scalar fields
$ \varphi_{1}^{\text{cl}} $, $ \varphi_{2}^{\text{cl}}$ and 
$\varphi_{3}^{\text{cl}}$ are zero. 
This procedure yields the specific functional form required by conformal invariance, as shown in Eq. \eqref{nonzeroT}, with
\begin{equation} \label{hweak}
{\mathbf h}^{(weak)}=-\frac{2}{g_{YM}^2} \frac{1}{12}\frac{(k^2-1)k}{8}=-\frac{1}{48 g_{YM}^2} (k^2-1)k \, .
\end{equation}

The weak coupling result of Eq. \eqref{hweak} matches the strong coupling expression in Eq. \eqref{hT}. To demonstrate this, the two results should be compared in the limit where
\begin{equation}\label{limit}
\frac{k}{\sqrt{\lambda}}=\frac{\kappa}{\pi}\gg 1 \quad 
\xRightarrow[]{\text{Eq.}  \eqref{induced-metric-limit1}} \quad
\sigma\rightarrow 0
\end{equation}
under which the coefficients ${\mathbf h}^{(weak)}$ and 
${\mathbf h}^{(strong)}$ reduce to
\begin{equation}\label{hweak-hstrong-agreement}
{\mathbf h}^{(weak)}= {\mathbf h}^{(strong)}=-\frac{\sqrt{2}}{24 g_{YM}^2}\frac{\lambda^{3/2}}{\pi^3 \sigma^3}\, .
\end{equation}
This provides a highly non-trivial test of the correspondence. A similar agreement was previously found in Refs. \refcite{Nagasaki:2012re} and \refcite{Kristjansen:2012tn} for CPOs in codimension-1 supersymmetric D3-D5 and non-supersymmetric D3-D7 systems, respectively. 
The logic here parallels the BMN limit, where a specific quantity—in this case $\lambda/{k^2}$, analogous to $\lambda/{J^2}$—is small at both weak and strong coupling, allowing for an order-by-order comparison of observables. For the supersymmetric codimension-2 D3-D3 defect, related studies can be found in Refs. \refcite{Gomis:2007fi} and \refcite{Drukker:2008wr}.
Before closing this section let us mention that from the value of ${\mathbf h}$ one can extract the anomaly coefficient $d_2$ that is associated with the pull-back of the Weyl tensor on the defect, see Ref. \refcite{Georgiou:2025mgg}. 
Indeed, the general relation connecting ${\mathbf h}$ and $d_2$ is, 
see Ref. \refcite{Lewkowycz:2014jia},
\begin{equation}
{\mathbf h} = -\frac{1}{2 \pi}\frac{1}{3\text{vol}(\mathbb{S}^{D-3})} \, \frac{D-3}{D-1} \, d_2 \, .
\end{equation}
Here $D$ is the number of dimensions of the ambient CFT in which the 
2-dimensional defect is embedded. In our case $D=4$.
Let us mention that the agreement between the weak and strong coupling results for ${\mathbf h}$ in the limit of Eq. \eqref{limit} directly implies a similar agreement between the corresponding values for the $d_2$ anomaly coefficient.

In addition, the anomaly coefficients $b$ and $d_1$ associated with the 
intrinsic and extrinsic curvatures of the defect for the dCFTs of 
Refs. \refcite{Georgiou:2025mgg} and \refcite{Georgiou:2025wbg} 
were recently calculated in Ref.~\refcite{Georgiou:2026tux}, 
at both the weak and strong coupling regimes. 
In a certain limit, agreement was found between the weak and strong coupling results. Notably, the anomaly coefficient $b$ was found to be negative for a finite range of the space of the parameters that characterize the dCFTs.


\subsection{One-point function of chiral primary operators}

In this section we derive the one-point functions for chiral primary operators (CPOs), which constitute a fundamental observable in defect CFTs. 
These functions vanish unless the CPO is invariant under the 
$SO(3)\times SO(3)$ symmetry group of the defect.
In accordance with Refs. \refcite{Nagasaki:2012re,Kristjansen:2012tn}, the spherical harmonic that preserves this symmetry is given by 
\begin{align} \label{sph}
Y_\Delta(\tilde\psi)  
&= \frac{    (2+\Delta)! }{  2^{(\Delta+1)/2} \sqrt{(\Delta+1)(\Delta+2)}  }
\sum_{p=0}^{\Delta/2} \frac{(-1)^p \sin^{\Delta-2p}\tilde\psi
\cos^{2p}\tilde\psi }{ (2p+1)!(1+\Delta-2p)! } \, . 
\end{align}


\subsubsection{Holographic computation}

The evaluation of the one-point function requires a perturbation of the Euclidean D5-brane action. To first order, the fluctuation is expressed as
\begin{equation} \label{perturbed-DBI-WZ}
S^{(1)} = \frac{T_5}{g_s} \int d^6 \zeta \left({\cal L}^{(1)}_{DBI}  +i\, 
{\cal L}^{(1)}_{WZ}\right) 
\end{equation}
where the Lagrangian components are defined by
\begin{equation} \label{Fluct_DBI+WZ}
{\cal L}^{(1)}_{DBI} = \frac{1}{2} \, \sqrt{\det H} \, \left(H^{-1}_{sym}\right)^{a b}
\partial_a X^M \, \partial_b X^N \, h_{MN}    \quad \& \quad 
\frac{{\cal L}^{(1)}_{WZ} }{2 \, \pi \alpha'}=\frac{F_{\beta \gamma} }{4!}\epsilon^{abcd} (Pa)_{abcd} \, . 
\end{equation}
Here, $h_{\mu \nu}$ and $a_4$ denote the fluctuations of the 
background metric and the RR 4-form, which are related to the supergravity field $s$ (dual to the CPO) via
\begin{align} \label{Fluct_components_metric_RR}
&h^\text{AdS}_{\mu\nu} 	=-\frac{2 \, \Delta \, (\Delta-1)}{\Delta+1}sg_{\mu\nu}+\frac{4}{\Delta+1}\nabla_\mu\nabla_\nu s
\nonumber \\[5pt]
&h^\text{S}_{\alpha\beta}	=2 \, \Delta \, s \, g_{\alpha\beta} 
\quad \& \quad
a^\text{AdS}_{\mu\nu\rho\sigma} = 4\, i \, 
\sqrt{g^\text{AdS}} \, \epsilon_{\mu\nu\rho\sigma\eta} \, \nabla^\eta s \, . 
\end{align}
The CPO is located at the point $(X_0,X_1,r',\psi',z=0)$ of the boundary 
with its location relative to the D5-brane illustrated 
in Fig. \ref{figg-2}. 
Subsequently, we substitute the expression for $s$ and integrate with respect to the six world-volume coordinates of the D5-brane. 
The DBI and the WZ contributions for a single CPO with conformal 
dimension $\Delta$ are
\begin{equation} \label{Full_DBI}
{\cal L}_{DBI} = +  \, \frac{1}{r'^{\Delta}}
\frac{2^{4-\Delta}\, \pi^3}{(1+\Delta) \,\sigma\,  \sqrt{8-\sigma^2}} 
\Bigg[ \Xi_1 \, \Psi_1 +\Xi_2\, \Psi_2+ \Xi_3  \, \Psi_3 \Bigg]
c_{\Delta}\, Y_{\Delta}(0) 
\end{equation}
\begin{equation} \label{Full_WZ}
{\cal L}_{WZ} = \,   \frac{1}{r'^{\Delta}}
\frac{2^{7-\Delta}\, \pi^3  \, (4+\sigma^2)}{(\Delta-1) \, \sigma^3 \, \sqrt{8-\sigma^2}} 
\Bigg[ \Xi_4 \, \Psi_2 +\left(\Delta - \Xi_4\right)\, \Psi_1 \Bigg] c_{\Delta}\, Y_{\Delta}(0) 
\end{equation}
where $P_{\delta}^{|k|}(x)$ are the associated Legendre polynomials, with $\Psi_1$, $\Psi_2$ and $\Psi_3$ we denote the following integrals\footnote{While the integrals $\Psi_1$, $\Psi_2$ and $\Psi_3$ 
do not admit closed-form solutions for arbitrary conformal dimensions, they can be explicitly evaluated for specific values of $\Delta$.}
\begin{equation} \label{Psi_1}
\Psi_1 = \int_0^{\frac{\pi}{2}} d \chi \, 
\sin^{\Delta -3} \chi \, 
P_{\Delta-2}^{(0)} \left(\sqrt{\big. 1+ \frac{1}{\sigma^2} \sin^2 \chi}\right)
\end{equation}
\begin{equation} \label{Psi_2}
\Psi_2 = \int_0^{\frac{\pi}{2}} d \chi \, 
\sqrt{\big.1+ \frac{1}{\sigma^2}\, \sin^2 \chi }\, 
\sin^{\Delta -3} \chi \, 
P_{\Delta-1}^{(0)} \left(\sqrt{\big. 1+ \frac{1}{\sigma^2} \sin^2 \chi}\right)
\end{equation}
\begin{equation} \label{Psi_3}
\Psi_3 = \int_0^{\frac{\pi}{2}} d \chi \, 
\Bigg[1+ \frac{1}{\sigma^2}\, \sin^2 \chi \Bigg]\, 
\sin^{\Delta -3} \chi \, 
P_{\Delta}^{(0)} \left(\sqrt{\big. 1+ \frac{1}{\sigma^2} \sin^2 \chi}\right) \, .
\end{equation}
and the values for the parameters $\Xi$ being 
\begin{eqnarray}
\Xi_1 &=& \frac{8 \left(5 \Delta ^2+\Delta -4\right) \sigma ^2+\Delta  (17 \Delta -15) 
\sigma ^4+32 \Delta  (\Delta +1)}{(1-\Delta ) \sigma ^2}
\nonumber \\[5pt] 
\Xi_2 &=& 32 \left(\Delta  \sigma ^2+\Delta +1\right) \, , 
\quad 
\Xi_3 =- \, 16 \, \Delta \,  \sigma ^2 \, , \quad 
\Xi_4 = \left(1- \Delta \right) \sigma^2 \, .
\end{eqnarray}
Substituting into Eq \eqref{perturbed-DBI-WZ} the DBI and the WZ contributions, we arrive to the following expression for the one-point function of the chiral primary operator
\begin{eqnarray}\label{O-general}
 \langle {\cal O}_\Delta(r')\rangle^{(strong)} &=&  \,    
 \frac{i^{\Delta } 2^{-\Delta -\frac{3}{2}}
   }{\pi \, \sqrt{\Delta }}\, \sqrt{\frac{\Delta +2}{\Delta +1}}\,
   \frac{\sigma }{\sqrt{8-\sigma ^2}} \,   \frac{\sqrt{\lambda }}{\Delta -1}  \, \frac{1}{r'^\Delta} \, 
\\
   &&\times  
   \Bigg[\left(9 \Delta ^2-15 \Delta +8\right)\,\Psi_1 -8 (\Delta -1) (3 \Delta -1) \Psi_2 +
   16 (\Delta -1) \Delta  \Psi_3\Bigg] \, .
   \nonumber
\end{eqnarray}
For $\Delta=4$ the one-point function becomes
\begin{eqnarray} \label{one-point-function-Delta-4}
  \langle {\cal O}_4(r')\rangle^{(strong)} & =  &  \frac{\sqrt{\lambda }}{16 \,  \sqrt{15} \, \pi } \, \frac{1}{r'^4}\,\frac{1+\sigma ^2}{\sigma ^5  \, \sqrt{8-\sigma ^2}}
   \, \left(5 \sigma ^4+56 \sigma
   ^2+96\right) \, . 
\end{eqnarray}


\subsubsection{Weak coupling computation \& agreement with holography}

Symmetry considerations for the D5-brane configuration imply that non-vanishing one-point functions are restricted to operators invariant under
$SO(3)\times SO(3)$ R-symmetry. 
A generic chiral primary operator of ${\cal N}=4$ SYM is defined as
\begin{equation}\label{chiralprimary}
{\cal O}_{\Delta I}(x)\equiv
\frac{(8\pi^2)^{\frac{\Delta}{2}}}
{\lambda^{\frac{\Delta}{2}}\sqrt{\Delta} }C_I^{i_1 i_2\ldots i_{\Delta}}
\,{\rm Tr}\left(\varphi_{i_1}(x)\varphi_{i_2}(x)\ldots \varphi_{i_{\Delta}}(x)\right),
\end{equation}
where $\varphi_i$ represent the six real scalar fields. 
The conformal dimension $\Delta$ remains independent of the 't Hooft coupling 
due to protection of supersymmetry. 
Furthermore, the tensors $C_{I}^{i_1i_2\ldots i_{\Delta}}$ are symmetric and traceless, satisfying the normalization condition
\begin{equation}\label{tracelesssymmetric}
\sum_{i_1\ldots i_\Delta=1}^6C_{I_1}^{i_1i_2\ldots i_{\Delta}}C_{I_2}^{i_1i_2\ldots i_{\Delta}}=\delta_{I_1I_2} \, .
\end{equation} 
Following the discussion in Ref. \refcite{Kristjansen:2012tn}, 
for every even conformal dimension $\Delta=2 l$, 
there exists a unique chiral primary operator, denoted by 
${\mathcal O}_\Delta(x')$, which is invariant under $SO(3)\times SO(3)$. 
By inserting the classical solution from Eq. \eqref{sol-2} into the trace in Eq. \eqref{chiralprimary}, we obtain the following result for the CPO one-point function
\begin{equation} \label{CPOweak}
\langle {\mathcal O}_\Delta(x')\rangle^{(weak)}=\frac{(8\pi^2)^{\frac{\Delta}{2}}}{\lambda^{\frac{\Delta}{2}}\sqrt{\Delta} }\frac{(k^2-1)^{\Delta/2}k}{(2 \sqrt{2} r')^\Delta}Y_\Delta(0) \, . 
\end{equation}

The results at weak and strong coupling coincide within the limit defined by Eq. \eqref{limit}. While we explicitly demonstrate this for the $\Delta=4$
case, we have confirmed that this consistency holds at least up to $\Delta=40$. By performing an expansion of Eqs. \eqref{one-point-function-Delta-4} and \eqref{CPOweak} near $\sigma =0$, we obtain
\begin{equation} \label{one-point-function-Delta-4-expansion}
  \langle {\cal O}_4(r')\rangle^{(weak)} = \langle {\cal O}_4(r')\rangle^{(strong)} = \sqrt{\frac{3}{10}} \, \frac{\sqrt{\lambda }}{\pi \, \sigma^5} \, \frac{1}{r'^4} \, +\cdots
\end{equation}
where the dots represent higher-order terms in the large 
$k/\sqrt{\lambda}\sim 1/\sigma$
expansion, which typically diverge between the two coupling regimes.

Furthermore, the leading-order contribution to the one-point functions 
of CPOs, in the limit of Eq. \eqref{limit}, can be generalized
to an arbitrary dimension $\Delta$ as
\begin{eqnarray}\label{genericD}
\langle {\cal O}_\Delta(r')\rangle^{(weak)} = 
\langle {\cal O}_\Delta(r')\rangle^{(strong)}& =  
(-1)^{\frac{\Delta}{2}} \, \sqrt{\frac{\Delta +2}{\Delta \left(\Delta +1\right)}} \, \frac{\sqrt{\lambda }}{\pi \, \sigma^{\Delta +1}} \, 
\frac{1}{r'^\Delta}  =  \nonumber \\
& =(-1)^{\frac{\Delta}{2}} \, \sqrt{\frac{\Delta +2}{\Delta \left(\Delta +1\right)}} \, \frac{\pi^{\Delta} \, k^{\Delta+1} }{2^{(\Delta+1)/2} \lambda^{\Delta/2}}  \, 
\frac{1}{r'^\Delta}.  
\end{eqnarray}
This general expression is derived by expanding the weak coupling result of Eq. \eqref{CPOweak} and applying the relation $k=\frac{\sqrt{2 \lambda}}{\pi \sigma}$. Our computations verify that Eq. \eqref{genericD} matches the corresponding strong coupling term in Eq. \eqref{O-general} for all tested dimensions up to $\Delta=40$.


\section{Conclusions}
\label{conclusions}

The purpose of this note is to review, in a concise and clear manner, recent developments in the holographic realisation of certain non-supersymmetric co-dimension two defect CFTs put forward in 
Refs.~\refcite{Georgiou:2025mgg} and \refcite{Georgiou:2025wbg}.

We begin by briefly discussing the co-dimension two 1/2-BPS configuration in its three complementary descriptions: as surface operators in gauge theory with singular field profiles, as probe branes in string theory, and as smooth ``bubbling'' supergravity geometries.

Subsequently, we focus on the holographic duality proposed in 
Ref.~\refcite{Georgiou:2025wbg}, which has the interesting feature of interpolating between a singular 1/2-BPS supersymmetric D5-brane probe at one end and the duality presented in Ref.~\refcite{Georgiou:2025mgg} at the other. 
For the latter, we review the calculation of several observables characterising the defect CFT, namely the one-point functions of the stress-energy tensor and of chiral primary operators. These calculations are performed at strong coupling using the probe-brane configuration and at weak coupling using the solution of the ${\mathcal N}=4$ SYM equations of motion conjectured to describe the defect.

In an appropriate limit, agreement between the weak- and strong-coupling results for these observables is found. This agreement constitutes a highly non-trivial test of the proposed duality since, as mentioned above, the presence of the defect completely breaks supersymmetry, leaving conformal symmetry as the only surviving symmetry. Consequently, the duality is far less constrained by symmetry than in the case of the 1/2-BPS Gukov--Witten operators. It is in this setting that the power of the holographic principle is most clearly manifested.

An important recent development is the evaluation of the defect Weyl anomaly coefficients $b$ and $d_1$, associated with the intrinsic and extrinsic curvature of the defect, respectively, performed in Ref.~\refcite{Georgiou:2026tux}. The agreement between the weak- and strong-coupling results for these coefficients, in the appropriate limit, provides further evidence for the validity and significance of the holographic dualities proposed in Refs.~\refcite{Georgiou:2025mgg} and \refcite{Georgiou:2025wbg}.

\section*{Acknowledgments}
This paper has been financed by the funding programme ``MEDICUS", of the University 
of Patras (D.Z. with grant number: 83800).

\section*{ORCID}

\noindent George Georgiou - 
\url{https://orcid.org/0000-0002-4848-4070}

\noindent Dimitrios Zoakos - \url{https://orcid.org/0000-0002-5991-4769}

\bibliographystyle{ws-ijmpd}

\bibliography{refs}

\end{document}